\documentclass{appolb}
\usepackage{graphicx}
\usepackage{amsmath}

\usepackage[square, numbers]{natbib}
\begin{document}
\title{Open and hidden heavy-ﬂavor hadron production in small systems with ALICE%
\thanks{Presented at Quark Matter 2022 - 29$^{th}$ international conference on ultrarelativistic nucleus-nucleus collisions}%
}
\author{Sébastien Perrin on behalf of the ALICE Collaboration
\address{Université Paris-Saclay Centre d’Etudes de Saclay (CEA), IRFU, Départment de Physique Nucléaire (DPhN),
Saclay, France}
\\
}
\maketitle
\begin{abstract}
Measurements of quarkonia (heavy quark and antiquark bound states) and open-heavy ﬂavour hadrons in hadronic collisions provide a unique testing ground for understanding quantum chromodynamics (QCD). Although recently there was signiﬁcant progress, our understanding of hadronic collisions has been challenged by the observation of intriguing effects in high-multiplicity proton---proton (pp) and p---Pb collisions, such as collective phenomena.
The excellent particle identiﬁcation, track and decay-vertex reconstruction capabilities of the ALICE experiment are exploited to measure quarkonia as well as open-heavy flavour hadron production at midrapidity while quarkonium measurements are inclusive at forward rapidity.
In this contribution, the ﬁrst measurements of the elliptic ﬂow ($v_2$) of $J/\psi$ in high multiplicity pp collisions is shown. New measurements of quarkonium and open-beauty hadron production in pp and p---Pb collisions are presented, such as the ﬁrst measurement of the non-prompt $\text{D}^{*+}$ polarization in pp collisions at $\sqrt{s}=13$ TeV. The comparison of results with available models is also discussed.

%Recently published inclusive quarkonium production cross sections at forward rapidity in pp collisions is presented as well. The comparison of results with available models is also discussed.

\end{abstract}
  
\section{Introduction}

The quark---gluon plasma (QGP) is a state of matter which is expected to be produced when the temperature and energy density are large enough for the hadronic matter to deconfine \cite{PBM_Quest}. Small systems, i.e. pp and p---Pb collisions, provide references to study production mechanisms in the vacuum and cold nuclear matter (CNM) effects. However, hints of collective behaviours, interpreted as a sign of deconfinement in A---A collisions, have been recently observed in high-multiplicity pp collisions \cite{CMS_ppflowoverall}. In order to investigate these behaviours, studies of dedicated observables like anisotropic flow as well as multiplicity-dependent studies, are carried out. In particular, the latter allow one to probe common behaviours across different system sizes as well as to shed light on multiparton interactions (MPIs), i.e. events in which two or more distinct parton-parton interactions occur simultaneously in a single hadron-hadron collision.
The use of open and hidden heavy-flavour probes is motivated by the unique insight they provide on the QGP phase. Indeed, heavy-flavour quarks are formed at the early stages of the collision due to the hard scale involved. In pp collisions, heavy flavour probes allow one to test perturbative QCD (pQCD) predictions and production mechanisms.

A detailed description of the ALICE apparatus can be found in Ref. \cite{ALICE_JINST}. In this section, we briefly present the detector systems relevant for the analyses discussed in these proceedings. At midrapidity, the tracking and determination of the interaction point is assured by the inner tracking system and its two innermost cylindrical layers, respectively. The time projection chamber (TPC) provides tracking and particle identification via the measurement of the specific energy loss ($\frac{dE}{dx}$) in the TPC gas. At forward rapidity, the minimum bias trigger and multiplicity measurement are assured by the V0 detector, an array of 32 scintillator counters and photo-multipliers spanning $-3.7<\eta<-1.7$ (V0C) and $2.8<\eta<5.1$ (V0A). The reconstruction of forward muons is assured by the muon spectrometer which covers $2.5<\eta<4.0$.

\section{Selected results in pp and p---Pb collisions}

\begin{figure}[htb]
\centerline{%
\includegraphics[width=\textwidth]{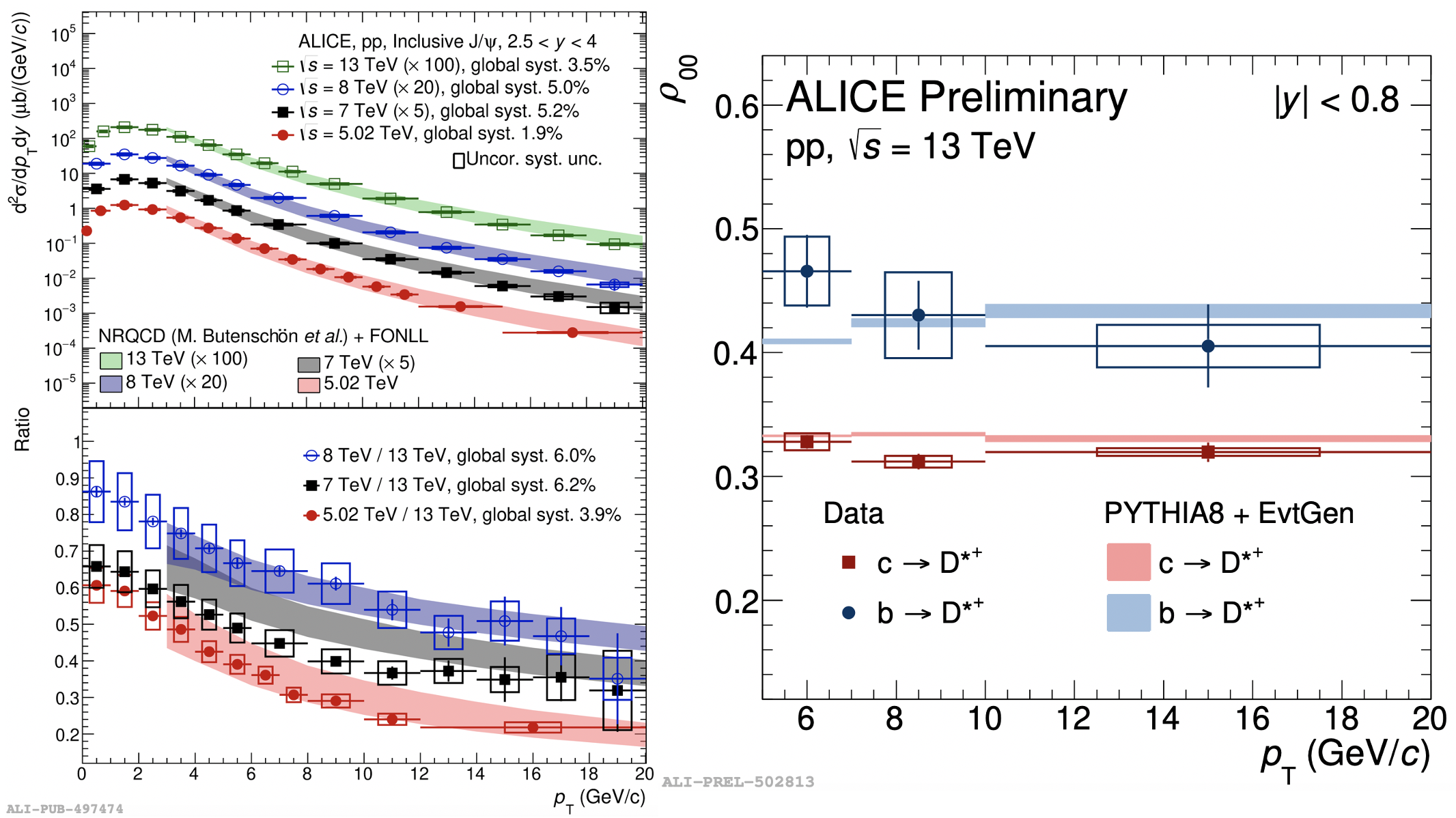}}
\caption{Left: Forward $J/\psi$ production cross section (top panel) and cross section ratios (bottom panel) for different energies and as a function of $p_\text{T}$\cite{ALICE_QaProd} compared to NRQCD + FONLL predictions \cite{NRQCD,FONLL}. Right: Measurement of the $\rho_{00}$ spin density matrix element for both prompt and non-prompt $\text{D}^{*+}$ as a function of $p_\text{T}$, compared to PYTHIA8+EVTGEN predictions \cite{PYTHIA,EVTGEN}.}
\label{Fig:Prod1}
\end{figure}

Using the entire pp data sample from Run 2, spanning centre-of-mass energies from 5.02 TeV to 13 TeV, ALICE measured the quarkonium production cross section at forward rapidity $2.5<y<4.0$ \cite{ALICE_QaProd}, providing measurements at $\sqrt{s_{\text{NN}}}=5.02$ TeV with ten times more statistics than the earlier publication \cite{ALICE_QaProd_Early}. In the left panel of Fig. \ref{Fig:Prod1}, the $J/\psi$ production cross section is presented as a function of transverse momentum ($p_\text{T}$) for different energies, showing a good agreement within uncertainties with NRQCD + FONLL predictions \cite{NRQCD,FONLL}. Cross section ratios impose additional constraints on models thanks to a partial cancellation of theoretical uncertainties. They evidence the difficulties the model has to reproduce simultaneously all the ratios among different energies, although data and model are still compatible within uncertainties. 

%In the right panel of Fig. \ref{Fig:Prod1}, the $p_\text{T}$-integrated cross section at forward rapidity is shown for various particle species as a function of the collision energy. Data from all species show a good agreement with the predictions from ICEM + FONLL \cite{ICEM, FONLL}, highlighting that quarkonium production is well described by theory over a large range of $p_\text{T}$, collision energies and particle species.

%Right: Forward quarkonium production cross sections as a function of collision energy compared to ICEM + FONLL predictions \cite{ICEM, FONLL}, from Ref. \cite{ALICE_QaProd}.

%\begin{figure}[htb]
%\centerline{%
%\includegraphics[width=6.5cm]{Plot2.png}}
%\caption{Measurement of the $\rho_{00}$ spin density matrix element for both prompt and non-prompt $\text{D}^{*+}$ with respect to $p_\text{T}$, compared to PYTHIA8+EVTGEN predictions \cite{PYTHIA,EVTGEN}.}
%\label{Fig:Polz2}
%\end{figure}

\noindent Heavy-flavour hadrons in small systems also serve as a baseline for larger systems which are expected to undergo hot nuclear effects. In Pb---Pb collisions, polarization measurements are expected to be impacted by the strong initial magnetic fields \cite{MagneticField} and large angular momentum arising from the non-zero impact parameter \cite{AngularMomentum}. The study of the $\text{D}^{*+}$ polarization was carried out at midrapidity ($|y|<0.8$) through the decay channel $\text{D}^{*+}\longrightarrow\text{K}^{-}\pi^{+}\pi^{+}$ in pp collisions at $\sqrt{s}=13$ TeV. The polarization is measured by studying the angular distribution of the decay products considering for the quantization axis the momentum direction of the vector meson $\text{D}^{*+}$ (helicity axis). It is parametrized by $\rho_{00}$, an element of the spin matrix $\rho$ which is 1/3 in case the particle is unpolarized. As the polarization of the $\text{D}^{*+}$ depends on its origin, both prompt (from direct production and excited state decays) and non-prompt (from beauty hadron decays) productions are studied and disentangled using machine learning techniques. In the right panel of Fig. \ref{Fig:Prod1}, the evolution of $\rho_{00}$ is shown as a function of $p_\text{T}$ for both prompt and non-prompt $\text{D}^{*+}$. The prompt $\text{D}^{*+}$ is unpolarized, whereas the non-prompt $\text{D}^{*+}$ shows a non-zero polarization. Both results are compatible with PYTHIA8 \cite{PYTHIA} predictions combined with EVTGEN \cite{EVTGEN}, the latter employed to simulate the $\text{D}^{*+}$ decay. This study demonstrates the ability to separate prompt and non-prompt sources using machine learning techniques to measure open beauty polarization, and gives a baseline for open beauty polarization measurements in larger systems.

In order to study collective effects in a specific collision system, the anisotropic flow observable \cite{Voloshin_Flow} is used. In a heavy-ion collision, the non-zero impact parameter makes the collision region anisotropic. This uneven geometry will result in anisotropies in the momentum distribution of the produced particles as the QGP expands. Measuring the angular distribution of the final state particles allows one to compute the flow coefficients ($v_n$). Each of them is sensitive to different effects (initial geometrical anisotropies for $v_2$, event-by-event fluctuations for $v_3$, etc.) and can then be compared to theory. 

\begin{figure}[htb]
\centerline{%
\includegraphics[width=7.5cm]{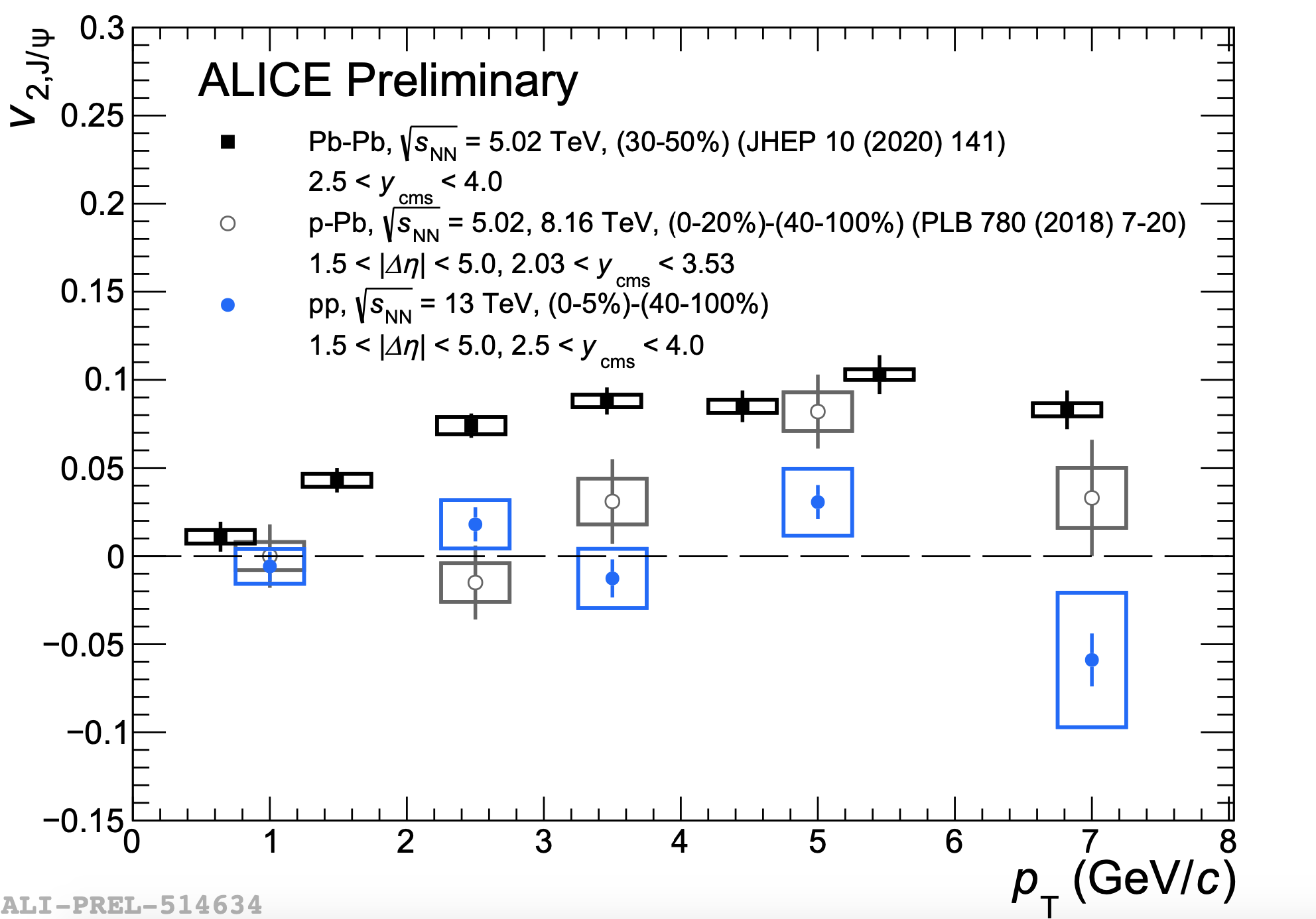}}
\caption{$J/\psi$ elliptic flow $v_{2}$ as a function of $p_\text{T}$ for pp, p---Pb \cite{Cvetan}, and Pb---Pb \cite{Robin} collision systems.}
\label{Fig:Flow3}
\end{figure}

\noindent The elliptic flow ($v_{2}$) has already been measured to be positive for light flavour particles at the LHC \cite{CMS_ppflowoverall}. For charm, ALICE concluded that the $J/\psi$ flows both in Pb---Pb \cite{Robin} and in p---Pb \cite{Cvetan} collisions with similar amplitudes. In Fig. \ref{Fig:Flow3}, the $J/\psi$ elliptic flow measured in pp collisions at $\sqrt{s}=13$ TeV at forward rapidity through the dimuon decay channel, is shown and compared to similar results from p---Pb\footnote{The p---Pb results are the average of $\sqrt{s_{\text{NN}}}=5.02$ TeV and $\sqrt{s_{\text{NN}}}=8.16$ TeV} and Pb---Pb collisions at $\sqrt{s_{\text{NN}}}=5.02$ TeV. A similar technique as the one employed in p---Pb collisions, based on the measurement of azimuthal correlations between inclusive $J/\psi$ and charged hadrons, is used. The results in pp collisions show no significant flow as a function of $p_\text{T}$ and the same conclusion holds also for the $p_\text{T}$-integrated case (1---12 GeV/$c$), where the deviation of $v_2$ from zero is about 1$\sigma$.

\begin{figure}[htb]
\centerline{%
\includegraphics[width=12.5cm]{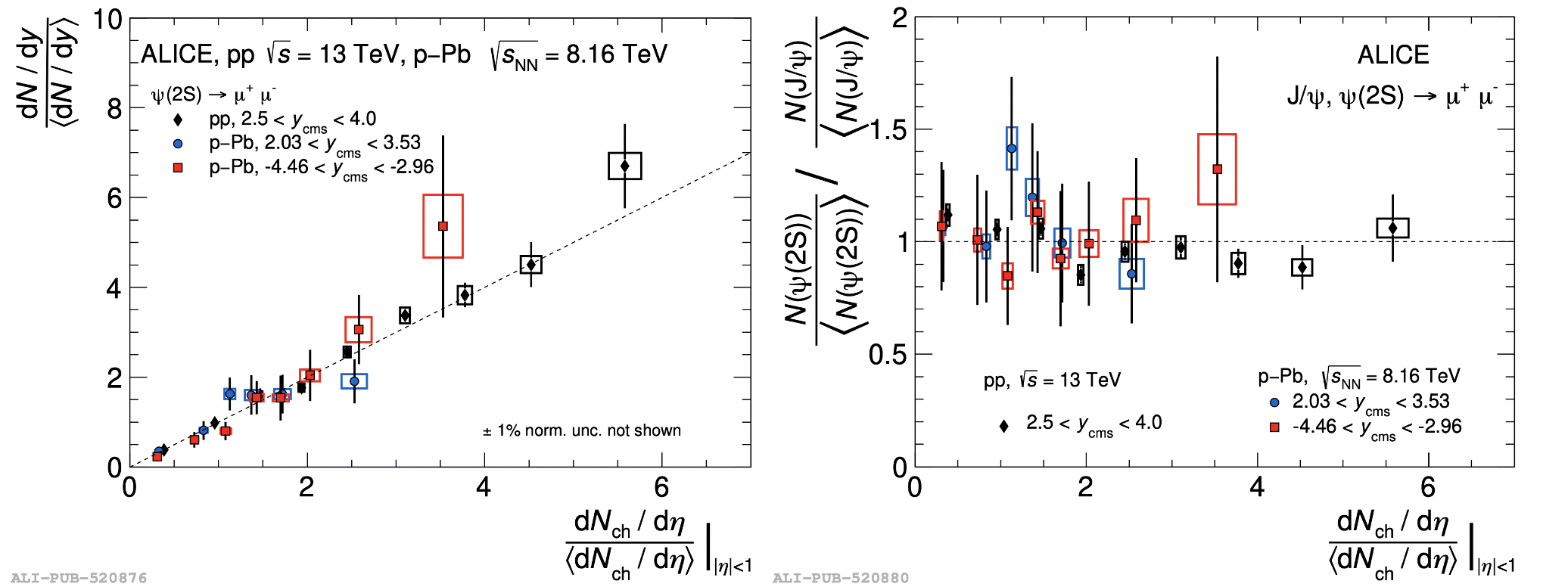}}
\caption{$\psi(2\text{S})$ self-normalized yield (left panel) and $\psi(2\text{S})$-to-$J/\psi$ self-normalized yield ratio (right panel) as a function of self-normalized charged particle pseudorapidity density in pp collisions at $\sqrt{s}=13$ TeV compared to the corresponding measurements in p---Pb and Pb---p collisions at $\sqrt{s_{\text{NN}}}=8.16$ TeV \cite{Theera}.}
\label{Fig:Prod4}
\end{figure}

%One of the possible explanations of collectivity in small systems are MPIs, which are accessible through multiplicity-dependent studies

\noindent While the origin of collective effects in small collision systems remains unclear, studying the production of the $\psi(2\text{S})$, in comparison to the more tightly bound $J/\psi$, could allow one to test the impact of final state effects, especially in high-multiplicity events where collective effects were evidenced for light flavours. In a new study by the ALICE collaboration \cite{Theera}, the quarkonium yields at forward rapidity are studied as a function of multiplicity. 
In the left panel of Fig. \ref{Fig:Prod4}, the self-normalized $\psi(2\text{S})$ yield measured at forward (pp, p---Pb) and backward rapidity (Pb---p) increases linearly as a function of the charged particle multiplicity density measured at midrapidity in pp collisions at $\sqrt{s}=13$ TeV, and in p---Pb and Pb---p collisions at $\sqrt{s_{\text{NN}}}=8.16$ TeV. In the right panel of Fig. \ref{Fig:Prod4}, the $\psi(2\text{S})$-to-$J/\psi$ yield ratio is shown in the same collision systems. For all systems, it is compatible with unity within uncertainties, pointing to a similar multiplicity dependence regardless of charmonium state and system size. In Ref. \cite{Theera}, the aforementioned multiplicity dependent results in pp collisions at $\sqrt{s}=13$ TeV are also compared with PYTHIA8 \cite{PYTHIA} and comovers model \cite{Comovers}. Both models manage to reproduce the self-normalized $\psi(2\text{S})$ yields as a function of charged particle multiplicity within uncertainties, while for the $\psi(2\text{S})$-to-$J/\psi$ yield ratio, some tensions with PYTHIA8 are visible at low multiplicity.

\section{Summary}

In these proceedings, some recent results of the ALICE collaboration concerning open and hidden heavy-flavour hadrons in small systems are presented. Quarkonium production cross sections at forward rapidity and prompt and non-prompt $\text{D}^{*+}$ polarization measurements at midrapidity are well described by available theory calculations within uncertainties. The first measurement of the $J/\psi$ elliptic flow in high-multiplicity pp collisions at $\sqrt{s}=13$ TeV does not show evidence for collective effects within current uncertainties. Multiplicity dependent results of charmonium production and excited-to-ground state ratios in pp and p---Pb collisions show weak dependence on the system size or charmonium state.

%uncomment the following lines to place a figure
%\begin{figure}[htb]
%\centerline{%
%\includegraphics[width=12.5cm]{Fig1}}
%\caption{Plot of ...}
%\label{Fig:F2H}
%\end{figure}

\bibliography{bibliography.bib}

\end{document}